\documentclass[aps,pra,preprint,groupedaddress]{revtex4-1}
\usepackage{graphicx}
\usepackage{amsmath}
\usepackage{amssymb}
\usepackage[hidelinks]{hyperref}
\usepackage{bm}

\newcommand{\ii}{\mathrm{i}}
\newcommand{\ee}{\mathrm{e}}

\begin{document}

\title{Ground state properties of interacting Bose polarons}

\author{Senne Van Loon}
\author{Wim Casteels}
\author{Jacques Tempere}
\affiliation{TQC, Universiteit Antwerpen, Universiteitsplein 1, 2610 Antwerpen, Belgium}

\date{\today}

\begin{abstract}
    We theoretically investigate the role of multiple impurity atoms on the
    ground state properties of Bose polarons. The Bogoliubov approximation is
    applied for the description of the condensate resulting in a Hamiltonian
    containing terms beyond the Fr\"ohlich approximation. The many-body nature
    of the impurity atoms is taken into account by extending the many-body
    description for multiple Fr\"ohlich polarons, revealing the static structure
    factor of the impurities as the key quantity. Within this formalism various
    experimentally accessible polaronic properties are calculated such as the
    energy and the effective mass. These results are examined for system
    parameters corresponding to two recent experimental realizations of the Bose
    polaron, one with fermionic impurities and one with bosonic impurities.  
\end{abstract}

\maketitle

\section{Introduction}
In general the complexity of many-body quantum systems makes it very hard to
make exact predictions starting from a microscopic model. In order to deal with
these systems one typically has to rely on approximations and an example with a
particularly appealing physical interpretation is the notion of quasiparticles.
They behave as effectively free particles but with properties that depend on the
many-body nature of the system. An example of such a quasiparticle which has
attracted a lot of interest over the years is the polaron. Its popularity can be
partly explained by its conceptual simplicity. The polaron was introduced by
Landau in 1933 for the description of an electron in a charged lattice
\cite{Landau}. In this picture the quasiparticle corresponds to the electron
together with its surrounding cloud of lattice polarization. Fr\"ohlich derived
a microscopic Hamiltonian for the polaron in terms of the electron interacting
with the lattice vibrations or phonons of the crystal
\cite{doi:10.1080/00018735400101213}. Various approximations and numerical
approaches have been applied to calculate the ground state properties of the
Fr\"ohlich Hamiltonian (see \cite{2010arXiv1012.4576D} for a detailed review). A
variational approach was developed by Lee, Low and Pines which is based on a
unitary transformation removing the electron degrees of freedom and which is
valid at weak and intermediate polaronic coupling \cite{PhysRev.90.297}. This
approach was later extended by Brosens, Lemmens, and Devreese in
Ref.~\cite{PSSB:PSSB2220820204} to examine the effect of multiple interacting
electrons which was incorporated through the static structure factor. 

More recently it has been realized that ultracold gases can be used as an
experimental platform to probe polaronic physics. Different classes of polarons
have been considered such as the Fermi polaron which corresponds to an impurity
atom in a quantum degenerate fermionic gas \cite{PhysRevLett.102.230402,
2012Natur.485..619K}. Another example, on which we will focus in this paper, is
an impurity atom immersed in a Bose condensed gas (see
Ref.~\cite{2015arXiv151004934G} for a review). This realization has led to a
revival of the Fr\"ohlich Hamiltonian since under certain conditions the same
polaron Hamiltonian applies as originally derived by Fr\"ohlich for the
description of an electron in a charged lattice \cite{PhysRevA.73.043608,
PhysRevLett.96.210401, PhysRevB.80.184504}. Within this mapping the role of the
electron is played by the impurity and the phonons are replaced by the
Bogoliubov excitations. This allowed to apply various theoretical approaches
that were originally developed for the solid state Fr\"ohlich polaron to the
Bose polaron. For example the extension of the Lee-Low-Pines approach for
multiple electrons was applied to multiple impurity atoms in a Bose-Einstein
condensate in Ref.~\cite{PhysRevA.84.063612} for fermionic impurities and in
Ref.~\cite{Casteels2013} for bosonic impurities. 

The applicability of the Fr\"ohlich Hamiltonian for the Bose polaron can only be
justified for weak interactions between the impurity and the bosons
\cite{PhysRevA.88.053610}. In recent years much effort has been devoted towards
the experimental realization of the Bose polaron \cite{PhysRevLett.103.170402,
PhysRevLett.107.135303, PhysRevLett.106.115305, PhysRevA.85.023623, Hohmann2015,
2016arXiv160501874R} and recently two set-ups have been presented that measure
the polaron energy and linewidth through RF-spectroscopy, one with fermionic
impurities \cite{PhysRevLett.117.055301} and one with bosonic impurities
\cite{PhysRevLett.117.055302}. These experiments apply a Feshbach resonance to
tune the impurity-boson scattering length, allowing to probe the system in the
whole range from weak to strong coupling. This revealed that the Fr\"ohlich
Hamiltonian is not adequate to describe all the different regimes, especially in
the regime of large impurity-boson scattering length. Considering certain terms
in the Hamiltonian that are neglected within the Fr\"ohlich approximation leads
to a much better agreement with the experimental observations
\cite{PhysRevLett.115.125302, PhysRevLett.117.113002,Grusdt2017,Brun2018Landau}.

In this paper we present a first step towards the description of a gas of
interacting Bose polarons beyond weak coupling. The descriptions that go beyond
the Fr\"ohlich approximation mostly consider just a single impurity. However in
the recent experimental set-ups presented in Refs.~\cite{PhysRevLett.117.055301,
PhysRevLett.117.055302} multiple impurities are present with an appreciable
density. We will extend the Lee-Low-Pines many-body approach to multiple
Fr\"ohlich polarons to the Hamiltonian that contains the additional beyond
Fr\"ohlich terms. We find that, as for the usual Fr\"ohlich polaron, the
structure factor of the impurities is the key ingredient needed to incorporate
the effect of multiple impurities. We use this to calculate experimentally
relevant properties such as the polaron energy and the effective mass. These
results are then examined for the system parameters corresponding to the
experiments of Refs.~\cite{PhysRevLett.117.055301, PhysRevLett.117.055302} and
the deviations from the single polaron results are discussed. 

The paper starts with introducing the Hamiltonian for the Bose polaron in
section \ref{Theory} and applying the Lee-Low-Pines approximation for multiple
impurities. In this section the analytical results for the polaron energy and
the effective mass are also presented. Then, in section \ref{ExpPar}, the two
recent experimental realizations of the Bose polaron are discussed with an
emphasis on the system parameters. Section \ref{Result} examines the results for
the parameters corresponding to these two experimental systems. Finally, in
section \ref{Conclusions}, we present the main conclusions and perspectives.

\section{Gas of Bose Polarons \label{Theory}}
We start from the microscopic Hamiltonian of a gas of bosons interacting with
impurities:
\begin{align}
    \hat{H} =  & \sum_{\bf{k}}\left(\epsilon_k -
    \mu\right)\hat{b}^{\dagger}_{\bf{k}}\hat{b}_{\bf{k}} +
    \frac{1}{2}\sum_{\bf{k},\bf{k}',\bf{q}}V_{BB}\left( \bf{q}\right)
    \hat{b}^{\dagger}_{\bf{k}'-\bf{q}} \hat{b}^{\dagger}_{\bf{k}+\bf{q}} 
    \hat{b}_{\bf{k}} \hat{b}_{\bf{k}'}
    + \sum_i^{N_I}\frac{\hat{p}^2_i}{2m_I} +
    \sum_{i<j}^{N_I}U\left(\hat{\bf{r}}_i - \hat{\bf{r}}_j\right) \nonumber \\
    &+ \sum_{\bf{k},\bf{q}}V_{IB}\left( \bf{q}\right) \hat{\rho}_{\bf{q}}
    \hat{b}^{\dagger}_{\bf{k}-\bf{q}}\hat{b}_{\bf{k}}
    \label{Ham}
\end{align}
The first two terms describe the bosons with chemical potential $\mu$, mass
$m_B$, kinetic energy $\epsilon_k = \hbar^2k^2/(2m_B)$ and
$V_{BB}\left(\bf{q}\right)$ the Fourier transform of the interaction amplitude.
The creation (annihilation) operators for the bosons are
$\{\hat{b}^{\dagger}_{\bf{k}}\}$ ($\{\hat{b}_{\bf{k}}\}$). The next two terms
describe the impurities with mass $m_I$ and momentum (position) operators
$\{\hat{\bf{p}}_i\}$ ($\{\hat{\bf{r}}_i\}$). The impurity-impurity interaction
potential is $U(r)$. The last term gives the interaction between the bosons and
the impurities with $V_{IB}\left( \bf{q}\right)$ the Fourier transform of the
interaction amplitude and $\hat{\rho}_{\bf{q}} = \sum_i^{N_I} e^{\ii
{\bf{q}}.\hat{{\bf{r}}}_i}$ the impurity density. For the sake of convenience we
will consider contact potentials for the impurity-boson and boson-boson
interaction potentials with strengths $g_{IB}$ and $g_{BB}$, respectively, i.e.
$V_{IB}\left( \bf{q}\right) = g_{IB}$ and $V_{BB}\left( \bf{q}\right) = g_{BB}$.
The temperature is considered sufficiently low such that the bosons form a
Bose-Einstein condensate which we describe with the Bogoliubov approximation for
weakly interacting bosons. The resulting Hamiltonian
\cite{PhysRevLett.117.113002} is
\begin{align}
    \hat{H}_\mathrm{Bog} =  & E_{GP} + g_{IB}N_I N_B +
    \sum_{\bf{k}}\hbar\omega_k\hat{\alpha}^{\dagger}_{\bf{k}}\hat{\alpha}_{\bf{k}}
    \nonumber \\
    & + g_{IB}\sqrt{N_0} \sum_{\bf{k}}W_{k} \hat{\rho}_{\bf{k}} 
    \left(\hat{\alpha}^{\dagger}_{-\bf{k}}
    + \hat{\alpha}_{\bf{k}}\right) \nonumber \\
    & +\frac{g_{IB}}{2} \sum_{{\bf{k}},{\bf{q}} \neq 0}\left( W_kW_q +
    W_k^{-1}W_q^{-1} \right)\hat{\rho}_{\bf{k} -
    \bf{q}}\hat{\alpha}^{\dagger}_{\bf{q}}\hat{\alpha}_{\bf{k}} \nonumber \\
    & +\frac{g_{IB}}{4} \sum_{{\bf{k}},{\bf{q}} \neq 0}\left( W_kW_q -
    W_k^{-1}W_q^{-1} \right)\hat{\rho}_{\bf{k} + \bf{q}}\left(
    \hat{\alpha}^{\dagger}_{-\bf{q}}\hat{\alpha}^{\dagger}_{-\bf{k}}
    +\hat{\alpha}_{\bf{q}}\hat{\alpha}_{\bf{k}} \right ) \nonumber \\
    &+ \sum_i^{N_I}\frac{\hat{p}^2_i}{2m_I} +
    \sum_{i<j}^{N_I}U\left(\hat{\bf{r}}_i - \hat{\bf{r}}_j\right)
    \label{HamBog}
\end{align} 
with $N_0$ the number of condensed bosons and $N_B$ the total number of bosons
within the Bogoliubov approximation, which will be approximated as $N_B \approx
N_0$.  $N_I$ is the number of impurities and we have also introduced the
function $W_k = \sqrt{\epsilon_k/(\hbar \omega_k)}$. Here $\omega_k$ is the
Bogoliubov dispersion $\omega_k = \hbar k\sqrt{(\xi k)^2+2}/(2m_B\xi)$, with
$\xi = \hbar/\sqrt{2 m_B N_0 g_{BB}}$ the condensate healing length. The
different terms on the first line of (\ref{HamBog}) are the mean field
Gross-Pitaevskii energy $E_{GP}$ of the condensate, the mean field interaction
energy and the energy of the Bogoliubov excitations that are created
(annihilated) by the operators $\{\hat{\alpha}^{\dagger}_{\bf{k}}\}$
($\{\hat{\alpha}_{\bf{k}}\}$). The second line represents the part of the
interaction between the impurity and the Bogoliubov excitations which is of the
Fr\"ohlich type. The third and the fourth lines are interaction terms which are
neglected within the Fr\"ohlich approximation. This is a good approximation if
the boson-impurity interaction strength is sufficiently weak. The last line in
(\ref{HamBog}) describes the kinetic energy and the mutual interactions of the
impurities, which is the same as in the original Hamiltonian (\ref{Ham}).

We will focus on the ground state properties of the Hamiltonian (\ref{HamBog}).
We do this by considering the many-body extension of the Lee-Low-Pines
transformation to the case of multiple interacting polarons. The corresponding
variational wave function is
\begin{equation}
    \left|\Psi \right> = \text{exp}\left[ \sum_{\bf{q}}
    \left(f_{\bf{q}}\hat{\alpha}_{\bf{q}}\hat{\rho}_{\bf{q}} -
    f^*_{\bf{q}}\hat{\alpha}^\dagger_{\bf{q}}\hat{\rho}_{-\bf{q}} \right)
    \right]\left|0 \right>\left|\psi \right>,
    \label{LLP}
\end{equation} 
with $\left|0 \right>$ the phonon vacuum, $\left|\psi \right>$ the wavefunction
of the impurities, and the $\{ f_{\bf{q}}\}$ are variational functionals. This
leads to the following variational expression for the energy:
\begin{align}
    E_\mathrm{Bog} =  & E_{GP} + g_{IB}N_I N_B +
    N_I\sum_{\bf{q}}|f_{\bf{q}}|^2\frac{\hbar^2q^2}{2m_I} +
    N_I\sum_{\bf{q}}\left[\hbar \omega_q|f_{\bf{q}}|^2 -
    g_{IB}\sqrt{N_0}W_{q}\left(f_{-\bf{q}} + f^*_{\bf{q}}\right)
    \right]S(\bf{q}) \nonumber \\
    & +\frac{g_{IB}}{2} \sum_{{\bf{k}},{\bf{q}} \neq 0}\left[\left( W_kW_q +
    W_k^{-1}W_q^{-1} \right)f^{}_{\bf{k}}f^*_{\bf{q}}+\frac{1}{2}\left( W_kW_q -
    W_k^{-1}W_q^{-1} \right)\left(f_{\bf{k}}f_{-\bf{q}} +
    f^*_{-\bf{k}}f^*_{\bf{q}} \right)\right] \nonumber \\
    & \kern6em \times \langle\psi|\hat{\rho}_{\bf{q} -
    \bf{k}}\hat{\rho}_{\bf{k}}\hat{\rho}_{-\bf{q}} |\psi\rangle +
    \sum_i^{N_I}\frac{\langle\hat{p}^2_i\rangle}{2m_I} + \sum_{i<j}^{N_I}\langle
    U\left( \hat{\bf{r}}_i - \hat{\bf{r}}_j\right)\rangle ,
    \label{EnVar}
\end{align}
where we introduced the structure factor $S(\bf{k})$ of the impurities
\begin{equation}
    S({\bf{k}}) = \frac{1}{N_I}
    \langle\psi|\hat{\rho}_{\bf{k}}\hat{\rho}_{-\bf{k}}|\psi\rangle .
    \label{SF}
\end{equation} 
In the variational expression for the energy (\ref{EnVar}) the interaction terms
that are beyond the Fr\"ohlich description lead to the presence of the anomalous
expectation value $\langle \psi|\hat{\rho}_{\bf{q} -
\bf{k}}\hat{\rho}_{\bf{k}}\hat{\rho}_{-\bf{q}} |\psi\rangle$. For a single
impurity this expectation value goes to one, i.e. $\lim_{N_I \to 1}
\langle\psi|\hat{\rho}_{\bf{q} -
\bf{k}}\hat{\rho}_{\bf{k}}\hat{\rho}_{-\bf{q}}|\psi \rangle = 1$, in which case
the variational polaron energy (\ref{EnVar}) can be minimized analytically
leading to an expression for the variational functionals $\{ f_{\bf{q}}\}$
\cite{PhysRevLett.117.113002}. This is however not possible in the general case
with a finite number of impurities and in order to proceed an approximation has
to be introduced. We note that the prefactor of the anomalous expectation value
$\langle\hat{\rho}_{\bf{q} - \bf{k}}\hat{\rho}_{\bf{k}}\hat{\rho}_{-\bf{q}}
\rangle$ is highly peaked around either ${\bf{k}} \to 0$ and ${\bf{q}} \to 0$.
In this limit we simply recover the impurity structure factor (\ref{SF}) and
with this motivation we introduce the following approximation:
\begin{equation}
\langle\hat{\rho}_{\bf{q}-\bf{k}}\hat{\rho}_{\bf{k}}\hat{\rho}_{-\bf{q}}\rangle
\rightarrow N_I S({\bf{k}}) S({\bf{q}}).
\label{AppSF}
\end{equation} 
A more detailed justification for this approximation can be found in the
Appendix. This allows us to minimize the energy (\ref{EnVar}) with respect to
the variational functionals $\{ f_{\bf{q}}\}$, leading to
\begin{equation}
    f_{\bf{k}} = \frac{2 \pi \hbar^2}{V m_r}\frac{\sqrt{N_0} W_{k}}{\hbar
    \omega_k + \frac{\hbar^2 k^2}{2m_I S(\bf{k})}}\left(a_{IB}^{-1} - a_0^{-1}
    \right)^{-1},
    \label{fq}
\end{equation} 
with $a_{IB}$ the impurity-boson scattering length which is related to the
interaction strength $g_{IB}$ through the Lippmann-Schwinger equation:
\begin{equation}
    \frac{V m_r}{2 \pi \hbar^2 a_{IB}} = g_{IB}^{-1} + \sum_{\bf{q}}\frac{2
    m_r}{\hbar^2 q^2},
\end{equation} 
with $m_r = m_I m_B/(m_I+m_B)$ the reduced mass. In expression (\ref{fq}) we
introduced the resonance length $a_0$ as
\begin{equation}
    a_0^{-1} = \frac{2 \pi \hbar^2}{V m_r}\sum_{{\bf{q}} \neq 0}\left(\frac{2
    m_r}{\hbar^2 q^2} - \frac{W_{q}^2S(\bf{q})}{\hbar\omega_q +
    \frac{\hbar^2 q^2}{2m_I S(\bf{q})}} \right).
    \label{ao}
\end{equation} 
Introducing the functionals (\ref{fq}) in the expression for the energy
(\ref{EnVar}) leads to
\begin{equation}
    \frac{E_\mathrm{Pol}}{N_I} = E_I + 
        \frac{2 \pi \hbar^2}{m_r}\frac{n_0}{a_{IB}^{-1} - a_0^{-1}},
    \label{PolEn}
\end{equation}
with 
\begin{equation}
    E_I = \frac{1}{N_I} \sum_i^{N_I} \frac{\langle\hat{p}_i^2\rangle}{2m} +
    \frac{1}{2}\sum_{{\bf{k}}}v({\bf{k}})[S({\bf{k}}) - 1].
\end{equation}
Here $n_0=N_0/V$ and $v(\bf{k})$ the Fourier transform of the impurity-impurity
interaction potential $U(r)$. Note that the resonant form of the polaron energy
(\ref{PolEn}) is similar to the expression found in
Ref.~\cite{PhysRevLett.117.113002}. The polaron effect leads to a shift of the
the impurity-boson resonance which is characterized by the resonance length
$a_0$. We stress that an important difference with the work in
Ref.~\cite{PhysRevLett.117.113002} is that now the resonance length $a_0$
depends on the many-body character of the impurities through the impurity static
structure factor. We note that the energy (\ref{PolEn}) does not exactly
converge to the single polaron result of Ref.~\cite{PhysRevLett.117.113002} in
the limit $N_I \to 1$. This is a consequence of neglecting phonon drag effects
in the current approximation. This was examined in
Ref.~\cite{PhysRevB.93.205144} within the Fr\"ohlich approximation, which
revealed that this is a very small effect that can be safely neglected. Based on
this we expect that also in the current case the inclusion of phonon drag
effects would not play an important role. An expansion of the polaron energy
(\ref{PolEn}) for small impurity-boson scattering length $a_{IB}$ up to second
order gives exactly the result derived in Ref.~\cite{PhysRevA.84.063612} within
the Fr\"ohlich approximation.

Another important characteristic of a quasiparticle in general is the effective
mass. The effect of the environment is then described by a quasiparticle with a
renormalized mass. This can be examined by allowing the impurities to move at a
constant speed $\bf{v}$. The total momentum of the system ${\bf{P}} =
\sum_i^{N_I}\hat{\bf{p}}_i + \sum_{\bf{k}}\hbar\bf{k} \,
\hat{\alpha}^{\dagger}_{\bf{k}}\hat{\alpha}_{\bf{k}}$ is a conserved quantity
and can thus be replaced by a number. The conservation of $\bf{P}$ can be made
explicitly by means of a Langrange multiplier $\bf{v}$ which physically
represents the velocity of the impurities. To describe the system we thus have
to minimize
\begin{equation}
    \hat{H}({\bf{v}}) = \hat{H}_\mathrm{Bog} -
    {\bf{v}}.\left(\sum_i^{N_I}\hat{\bf{p}}_i + \sum_{\bf{k}}\hbar\bf{k} \,
    \hat{\alpha}^{\dagger}_{\bf{k}}\hat{\alpha}_{\bf{k}} - \hat{\bf{P}} \right).
    \label{PolEn2}
\end{equation}
We can now follow the same steps as above leading to the following expression
for the variational functionals:
\begin{equation}
    f_{\bf{k}} = \frac{2 \pi \hbar^2}{V m_r}\frac{\sqrt{N_0} W_{k}}{\hbar
    \omega_k - \hbar {\bf{v}}.{\bf{k}} + \frac{\hbar^2 k^2}{2m_I S({\bf{k}})}}
    \left(a_{IB}^{-1} - a_0^{-1}  \right)^{-1}.
    \label{fq2}
\end{equation} 
Minimization of (\ref{PolEn2}) with respect to $\bf{v}$ leads to a relation
between the momentum $\bf{P}$ and the velocity $\bf{v}$ which in the limit of
small $|\bf{v}|$ can be written as ${\bf{P}} = N_Im^*\bf{v}$, with $m^*$ the
effective mass:
\begin{equation}
    m^* = m_I + \frac{2}{3} \frac{4 \pi^2 \hbar^4}{V^2
    m_r^2}\frac{1}{\left(a_{IB}^{-1} - a_0^{-1} \right)^2} \sum_{\bf{q}}\hbar^2
    q^2 \frac{N_0 W_q^2 S({\bf{q}})^4}{\left(S({\bf{q}})\hbar \omega_q +
    \frac{\hbar^2 q^2}{2m_I} \right)^3} .
    \label{meff}
\end{equation}
Note that we considered the system to be three dimensional and homogeneous. The
leading order contribution in an expansion for small $a_{IB}$ again gives the
result derived in \cite{PhysRevA.84.063612} with the Fr\"ohlich Hamiltonian, as
expected. Note that in the expression for the effective mass in
Ref.~\cite{PhysRevA.84.063612} the structure factor should be raised to the
fourth power in stead of being squared.

This reveals that, as for the Fr\"ohlich polaron, the influence of multiple
impurities is captured by the static structure factor of the impurities. For the
wave function of the impurities $\left|\psi \right>$ we introduce the
unperturbed wavefunction, neglecting the presence of the bosons. For fermionic
impurities at low temperature the impurity-impurity interactions can be
neglected and the structure factor for a free gas of fermions is
\begin{equation}
    S_F(\bf{k}) = \begin{cases} \frac{3}{2}\frac{k}{2k_F} -
    \frac{1}{2}\left(\frac{k}{2k_F}\right)^3 &\text{  for   } k<2k_F  \\
    1 &\text{  for   } k\geq2k_F
    \end{cases}
    \label{SFFer}
\end{equation}
where $k_F = (6 \pi^2 n_I)^{1/3}$ is the Fermi wavevector and $n_I$ the impurity
density. For bosonic impurities we assume them to be weakly interacting and to
form a condensate such that we can calculate the dynamic structure factor within
the Bogoliubov approximation, resulting in
\begin{equation}
    S_B({\bf{k}}) = \frac{k}{\sqrt{k^2 + 16 \pi n_I a_{II}}},
    \label{SFBos}
\end{equation}
with $a_{II}$ the impurity-impurity scattering length and $n_I$ the impurity
density. 

In order to derive the results in this section the approximation (\ref{AppSF})
was introduced. We would like to stress that the main motivation for this
approximation is that it allows to calculate analytically the polaronic
properties, while capturing the many-body nature of the impurities to a certain
extent. In this way, we approximate the three-point correlations as a small
correction to the two-point correlations. As also stated in the introduction we
consider this a first step towards the description of a gas of interacting Bose
polarons.

\section{Experimental parameters \label{ExpPar}}
In the next section we will examine our results in the context of two recent
experiments where the polaron RF-spectrum was measured
\cite{PhysRevLett.117.055301,PhysRevLett.117.055302}. In this section we briefly
discuss the system parameters for these two experiments.

A recent experiment on the Bose polaron was performed at JILA and presented in
Ref.~\cite{PhysRevLett.117.055301} using $^{87}\text{Rb}$ for the condensed
bosons and fermionic $^{40}\text{K}$ impurities. The boson-boson scattering
length is $a_{BB} = 100 \, a_{\text{Bohr}}$ and the peak boson density is $n_0 =
1.8 \times 10^{14} \,\text{cm}^{-3}$. This leads to a condensate healing length
$\xi = 200 \, \text{nm}$ and a gas parameter $n_0^{1/3}a_{BB} \approx 0.3$. The
density of the fermionic impurities is $n_I = 2 \times 10^{12} \,
\text{cm}^{-3}$ leading to a Fermi wavevector $k_F = 5 \, \mu\text{m}^{-1}$. In
units of the healing length the impurity density is thus $n_I\xi^3 = 0.016$. 

Another experimental realization of the Bose polaron was performed at the
university of Aarhus, Denmark and is presented in
Ref.~\cite{PhysRevLett.117.055302}. In their set-up both the condensed bosons
and the impurities are different internal states of the same bosonic
$^{39}\text{K}$ atom. The condensate density is $n_0 = 2.3 \times 10^{14} \,
\text{cm}^{-3}$ and the boson-boson scattering length is $a_{BB} = 9 \,
a_{\text{Bohr}}$ which gives a healing length $\xi = 200 \, \text{nm}$ and a gas
parameter $n_0^{1/3}a_{BB} \approx 0.03$. The fraction of excited impurities is
of the order of $10\%$.

\section{Results \label{Result}}
\begin{figure}[t!]
    \includegraphics[scale=1]{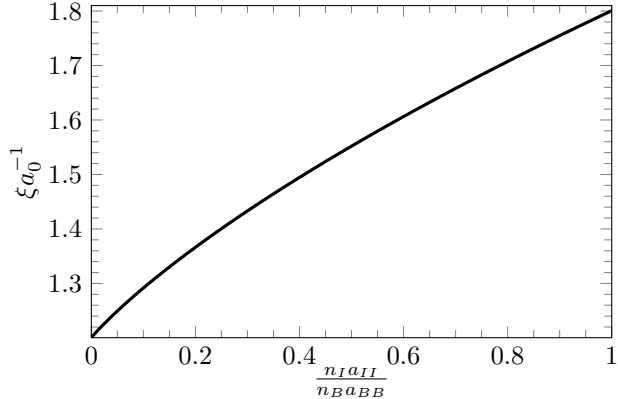}
    \caption{\label{Fig1} 
    The inverse resonance length $a_0^{-1}$ (units of the inverse healing length
    $\xi$) for bosonic impurities as a function of $n_Ia_{II}/(n_0 a_{BB})$,
    i.e. the ratio of the densities times the ratio of the scattering lengths.
    The masses of the impurities and the bosons are considered the same, i.e.
    $m_B = m_I$ (corresponding to the experimental realization of
    Ref.~\cite{PhysRevLett.117.055302}). }
\end{figure}

\begin{figure}[t!]
    \includegraphics[scale=1]{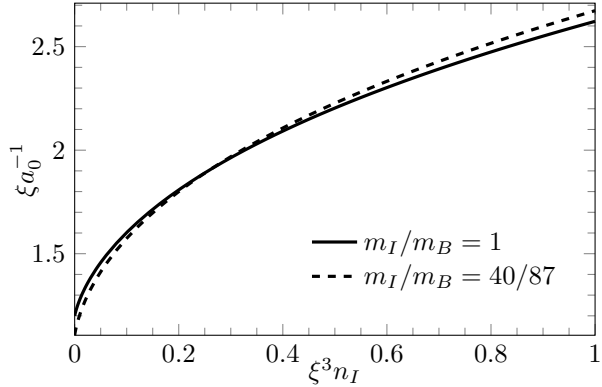}
    \caption{\label{Fig2} 
    The inverse resonance length $a_0^{-1}$ (units of the inverse healing length
    $\xi$) as a function of the impurity density $n_I$ (units of $\xi^{-3}$) for
    fermionic impurities. The full line is for equal impurity and boson masses,
    i.e. $m_B = m_I$, the dashed line corresponds to the mass imbalance for a
    $^{40}\text{K}$ impurity in a $^{87}\text{Rb}$ condensate, i.e. $m_I/m_B =
    40/87$ (corresponding to the experimental realization of
    Ref.~\cite{PhysRevLett.117.055301}).}
\end{figure}

In this section we examine the results derived in Section \ref{Theory} for the
experimental system parameters discussed in Section \ref{ExpPar}. In
Figs.~\ref{Fig1} and \ref{Fig2} the inverse resonance length $a_0^{-1}$ is
presented as a function of the impurity density $n_I$ for the bosonic and the
fermionic impurities, respectively. In both cases the inverse resonance length
increases as a function of the impurity density and in the many-impurity limit
$n_I \to \infty$ the resonance length vanishes, i.e. $a_0 \to 0$ corresponding
to the disappearance of the polaronic effect. This behavior was also found
within the Fr\"ohlich approximation in Refs.~\cite{PhysRevA.84.063612,
Casteels2013}. In Fig.~\ref{Fig2} the resonance length is presented both for the
mass balanced case ($m_I = m_B$, full line) and for the experimental system
discussed in Section \ref{ExpPar} with a mass imbalance ($m_I/m_B = 40/87$,
dashed curve).

\begin{figure}[t!]
    \includegraphics[scale=1]{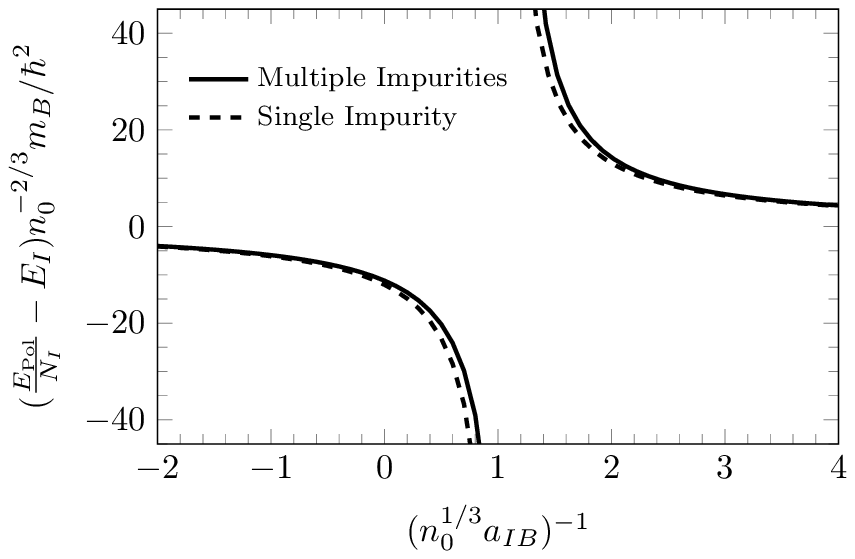}
    \caption{\label{Fig3} 
    The polaron energy as a function of the inverse boson-impurity scattering
    length $a_{IB}$ (in units of $n_0^{1/3}$) for bosonic impurities in a
    Bose-Einstein condensate. The dashed line is for a single impurity while the
    full curve corresponds to a non-zero impurity density characterized by
    $n_Ia_{II}/(n_0a_{BB}) = 0.1$. The condensate gas parameter is set to
    $n_0^{1/3}a_{BB} = 0.03$ and the impurity and boson masses are the same,
    i.e. $m_B = m_I$ (corresponding to the experimental realization presented in
    Ref.~\cite{PhysRevLett.117.055302}). }
\end{figure}

\begin{figure}[t!]
    \includegraphics[scale=1]{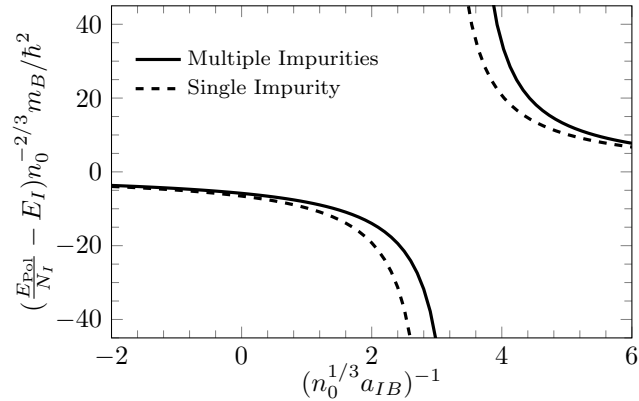}
    \caption{\label{Fig4} 
    The polaron energy as a function of the inverse boson-impurity scattering
    length $a_{IB}$ (in units of $n_0^{1/3}$) for fermionic impurities in a
    Bose-Einstein condensate. The dashed line is for a single impurity while the
    full curve corresponds to a non-zero impurity density characterized by
    $n_I\xi^3 = 0.016$. The condensate gas parameter is set to $n_0^{1/3}a_{BB}
    = 0.3$ and the impurity-boson mass imbalance is $m_I/m_B = 40/87$
    (corresponding to the experimental realization presented in
    Ref.~\cite{PhysRevLett.117.055301}). }
\end{figure}

In Figs.~\ref{Fig3} and \ref{Fig4} the polaron energy is presented as a function
of the inverse boson-impurity scattering length $a_{IB}^{-1}$. This reveals the
well-known resonance behavior with an attractive and a repulsive polaron branch
(as also described in Refs.~\cite{PhysRevA.88.053632, PhysRevLett.117.055302,
PhysRevLett.117.113002} for a single impurity). In Fig.~\ref{Fig3} the result
for a single impurity is compared with the result in the presence of a
Bose-condensed gas of bosonic impurities characterized by the ratio
$n_Ia_{II}/(n_0a_{BB}) = 0.1$. This clearly reveals a shift of the resonance
position due to the presence of multiple interacting impurities. In
Fig.~\ref{Fig4} a similar shift of the resonance is found for the case of
non-interacting fermionic impurities characterized by the dimensionless quantity
$n_I\xi^3 = 0.016$ and with a mass imbalance $m_I/m_B = 40/87$.

\begin{figure}[t!]
    \includegraphics[scale=1]{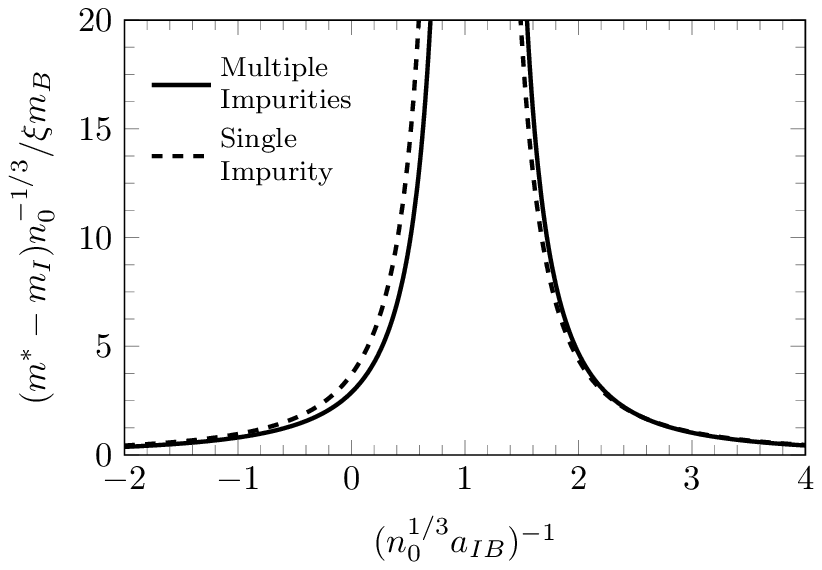}
    \caption{\label{Fig5} 
    The polaron effective mass as a function of $(n_0^{1/3}a_{IB})^{-1}$ for
    bosonic impurities. The condensate gas parameter is taken $n_0^{1/3}a_{BB} =
    0.03$ and the masses of the impurity and the bosons are considered equal,
    i.e. $m_B = m_I$. The dashed line is for a single impurity while the full
    curve corresponds to an impurity density of $n_Ia_{II}/(n_0a_{BB}) = 0.1$
    (corresponding to the experimental realization presented in
    Ref.~\cite{PhysRevLett.117.055302}).}
\end{figure}

\begin{figure}[t!]
    \includegraphics[scale=1]{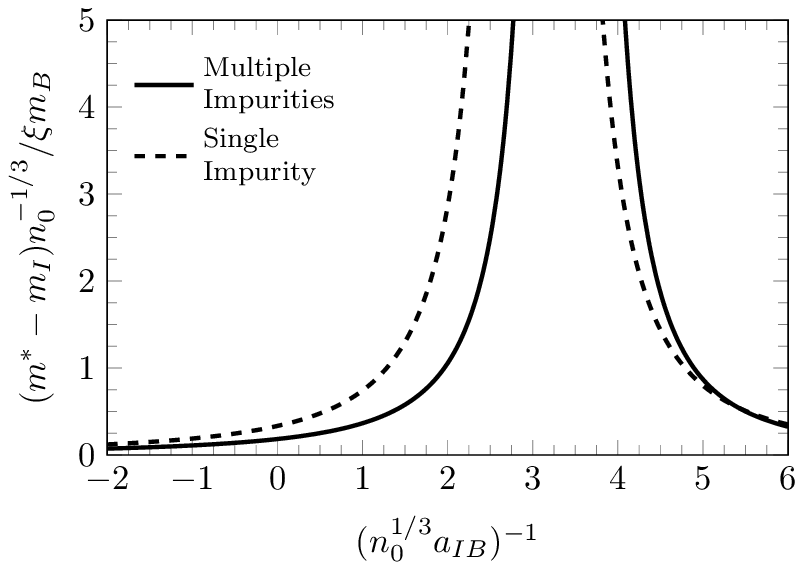}
    \caption{\label{Fig6} 
    The polaron effective mass as a function of $n_0^{1/3}a_{IB}$ for fermionic
    impurities. The condensate gas parameter is taken $n_0^{1/3}a_{BB} = 0.3$
    and an impurity-boson mass imbalance of $m_I/m_B = 40/87$ is considered. The
    dashed line is for a single impurity while the full curve corresponds to a
    finite impurity density characterized by $n_I\xi^3 = 0.016$. (corresponding
    to the experimental realization presented in
    Ref.~\cite{PhysRevLett.117.055301}). }
\end{figure}

In Figs.~\ref{Fig5} and \ref{Fig6} the effective mass is presented as a function
of the inverse impurity-boson scattering length $a_{IB}^{-1}$ for bosonic and
fermionic impurities, respectively. This reveals that the effective mass
increases as the resonance is approached. At the resonance the effective mass
diverges, signaling the break-down of the polaron quasiparticle picture in this
regime, in agreement with the conclusions of Ref.~\cite{PhysRevLett.117.113002}.
Similar as for the energy we again observe a clear shift of the resonance due to
the many-body nature of the impurities, both for the bosonic and the fermionic
impurities.

\section{Conclusions and perspectives \label{Conclusions}}
We have extended the many-body description for a gas of Fr\"ohlich polarons to
interacting Bose polarons, taking into account terms in the Hamiltonian that are
beyond the Fr\"ohlich approximation. In order to solve the resulting equations
analytically we have introduced an approximation for the anomalous expectation
value containing three times the impurity density operator. As for the
Fr\"ohlich polaron, the key ingredient that characterizes the impurity gas is
the static structure factor. This approximation also allowed us to derive the
effective mass of the Bose polarons. These results were applied for the system
parameters corresponding to two recent experimental realizations of the Bose
polaron \cite{PhysRevLett.117.055301,PhysRevLett.117.055302}. This reveals a
shift of the resonance position with respect to the single polaron result, which
is a consequence of the many-body nature of the impurities. Our results also
show that in the limit of many impurities the polaronic mass effect disappears.
As the resonance is approached the polaron effective mass increases and
ultimately diverges which indicates the break-down of our description close to
the resonance. A future perspective is about the fate of the Bose polaron in
this regime of strong impurity-boson interaction close to the resonance. An
extension of the polaronic strong coupling theory, which is well-established for
the Fr\"ohlich polaron \cite{PhysRevB.46.301, PhysRevA.73.043608,
PhysRevLett.96.210401, Casteels2011}, to the Bose polaron could shine light on
this interesting open question. 

\acknowledgements{The authors gratefully acknowledge discussions with J.T.
    Devreese, G. Lombardi, J. Arlt and G. Bruun . This work was supported by the
    joint FWO-FWF project POLOX (Grant No.  I 2460-N36), by the Fund for
    Scientific Research – Flanders grant nr. G.0429.15.N and by the Research
    Council of Antwerp University.}

\appendix*

\section{Anomalous expectation value}

In this Appendix we clarify the approximation \eqref{AppSF} of the anomalous
expectation value
\begin{equation}
    \mathcal{I}({\bf{k}},{\bf{q}}) = \langle \psi|\hat{\rho}_{{\bf{q}} - {\bf{k}}}
    \hat{\rho}_{{\bf{k}}}\hat{\rho}_{-{\bf{q}}} |\psi\rangle = 
    \sum\limits_{j,m,n=1}^{N_I} \langle 
        \ee^{\ii {\bf{q}}.(\hat{{\bf{r}}}_j - \hat{{\bf{r}}}_n)}
        \ee^{\ii {\bf{k}}.(\hat{{\bf{r}}}_m - \hat{{\bf{r}}}_j)} \rangle.
    \label{eq:AnomExp}
\end{equation}
This expectation value is a special case of the three-point correlation function
$\langle \hat{\rho}_{\bf{k_1}} \hat{\rho}_{\bf{k_2}} \hat{\rho}_{\bf{k_3}}
\rangle$. The density operators can be expanded around their expectation value
$\hat{\rho}_{\bf{k}} \rightarrow \hat{\rho}_{\bf{k}} - \langle
\hat{\rho}_{\bf{k}} \rangle,$ which connects the moments to the central moments,
as one considers the three-point correlation function as the third moment of the
density distribution. Then approximating that the third central moment is zero
provides an approximation of the three-point correlation function in terms of
one- and two-point correlation functions. Performing the central moment
approximation to the two-point correlation function corresponds to the Hartree
approximation, which would reduce the three-point correlation function to a
product of three expectation values of the density.

We extend this idea to the anomalous expectation value \eqref{eq:AnomExp} by
introducing the structure factor \eqref{SF} to write down the central moment 
\begin{equation}
\mathcal{I}_\mathrm{central}({\bf{k}},{\bf{q}})=\sum\limits_{j,m,n=1}^{N_I}
\left\langle 
    \left( \ee^{\ii {\bf{q}}.(\hat{{\bf{r}}}_j - \hat{{\bf{r}}}_n)} 
        - \frac{1}{N_I} S({\bf{q}}) \right)
    \left( \ee^{\ii {\bf{k}}.(\hat{{\bf{r}}}_m - \hat{{\bf{r}}}_j)} 
        - \frac{1}{N_I} S({\bf{k}}) \right) \right\rangle .
\end{equation}
The approximation $\mathcal{I}_\mathrm{central} \approx 0$ then reduces to
\begin{equation}
    \mathcal{I}({\bf{k}},{\bf{q}}) \approx N_I S({\bf{k}}) S({\bf{q}}),
    \label{eq:Iappr}
\end{equation}
as in Eq.~\eqref{AppSF}. Note that when $\bf{q} \to 0$ ($\bf{k} \to 0$), we get
$\mathcal{I}({\bf{k}},0) = N_I^2 S({\bf{k}})$ (and $\mathcal{I}(0,{\bf{q}}) =
N_I^2 S({\bf{q}})$), such that the central moment approximation is exact in
these limits (since $S(0) = N_I$). In the expression for the polaronic energy
\eqref{EnVar} the terms where $\mathcal{I}({\bf{k}},{\bf{q}})$ occurs have to be
summed over all $\bf{k},\bf{q}$ after multiplication by a prefactor. As this
prefactor is largest when either $\bf{k}$ or $\bf{q}$ are close to zero, the
terms where the approximation \eqref{eq:Iappr} holds can be expected to provide
the dominant contribution.

\bibliography{manusc}

\end{document}